\documentclass[12pt,preprint]{aastex}

\slugcomment{\today}

\shorttitle{Quarks and Pions in Black Hole-Forming Stellar Collapse}
\shortauthors{Nakazato, Sumiyoshi, \& Yamada}

\begin{document}


\title{Impact of Quarks and Pions on Dynamics and Neutrino Signal of \\ Black Hole Formation in Non-rotating Stellar Core Collapse}


\author{Ken'ichiro Nakazato\altaffilmark{1}, Kohsuke Sumiyoshi\altaffilmark{2} and Shoichi Yamada\altaffilmark{3,4}}

\email{nakazato@kusastro.kyoto-u.ac.jp}


\altaffiltext{1}{Department of Astronomy, Kyoto University, Kita-shirakawa Oiwake-cho, Sakyo, Kyoto 606-8502, Japan}
\altaffiltext{2}{Numazu College of Technology, Ooka 3600, Numazu, Shizuoka 410-8501, Japan}
\altaffiltext{3}{Department of Physics, Waseda University, 3-4-1 Okubo, Shinjuku, Tokyo 169-8555, Japan}
\altaffiltext{4}{Advanced Research Institute for Science and Engineering, Waseda University, 3-4-1 Okubo, Shinjuku, Tokyo 169-8555, Japan}


\begin{abstract}
In the formation process of black holes, the density and temperature of matter become sufficiently high for quarks and pions to appear. In this study we numerically investigate stellar core collapse and black hole formation taking into account the equations of state involving quarks and/or pions. In our simulations, we utilize a code that solves the general relativistic hydrodynamics and neutrino transfer equations simultaneously, treating neutrino reactions in detail under spherical symmetry. Initial models with three different masses, namely, 40, 100 and $375M_\odot$, are adopted. Our results show that quarks and pions shorten the duration of neutrino emission if the collapse bounces before black hole formation. In addition, pions increase the luminosity and average energy of neutrinos before black hole formation. We also find that the hadron-quark phase transition leads to an interesting evolution of temperature. Moreover, the neutrino event number is evaluated for the currently operating neutrino detector, SuperKamiokande, to confirm that it is not only detectable but also affected by the emergence of quarks and pions for Galactic events. While there are some issues, such as hyperons, beyond the scope of this study, this is the first serious attempt to assess the impact of quarks and pions in dynamical simulations of black hole formation and will serve as an important foundation for future studies. 
\end{abstract}





\section{Introduction} \label{intro}

Massive stars with the main-sequence mass $M\gtrsim 10M_\odot$ are known to undergo gravitational collapse at the end of their lives \citep[][]{poe08}. In particular, stars with $M\lesssim 25M_\odot$ are thought to end their lives as type II supernovae. It is thought that the accompanying explosion is invoked by the shock wave launched by the core bounce due to the nuclear repulsion, leaving a neutron star. On the other hand, the fate of stars with $M\gtrsim 25M_\odot$ can be observationally split into two branches, namely, a hypernova branch and a faint-supernova branch, and they are both thought to form black holes eventually \citep[][]{nomoto06}. It has been proposed that strongly rotating massive stars are constituents of the hypernova branch while nonrotating and weakly rotating massive stars are constituents of the faint-supernova branch. Nonrotating stars more massive than faint-supernova progenitors are thought to result in so-called failed supernovae, where the shock produced by the bounce cannot propagate outward owing to the thick outer layer. In fact, a recently discovered black hole candidate with 24-$33M_\odot$ \citep[][]{pres07,silfil08} may be a remnant of a failed supernova, and a survey involving the monitoring of $\sim \!\! 10^6$ supergiants has been proposed to investigate the end of lives of such massive stars \citep[][]{kocha08}. It should be noted that the hypothesis described above is a hot subject under active discussion.

So far, many numerical simulations of the gravitational collapse of massive stars have been performed in an attempt to elucidate the mechanism of core-collapse supernovae \citep[e.g.,][]{colgate66,totani98,marek09}, although definitive results are still lacking. Numerical studies on black hole formation by stellar collapse have also begun to be carried out. \citet{fryer99} showed that nonrotating stars with $25M_\odot \lesssim M\lesssim 40M_\odot$ produce a faint-supernova explosion. In this case, a proto--neutron star is formed, which is thought to recollapse to a black hole $\gtrsim \!\! 10$~s after the bounce \citep[e.g.,][]{baum96}. On the other hand, nonrotating stars with $M\gtrsim 40M_\odot$ become failed supernovae involving prompt ($\sim \!\! 1$~s after the bounce) black hole formation \citep[e.g.,][]{sumi07,sumi08,fischer09}. The gravitational collapse of massive stars is accompanied by the emission of a large amount of neutrinos. In the case of SN1987A, the emitted neutrinos were actually detected \citep[][]{hirata87,bionta87}. Neutrinos are thought to play an essential role in the explosion mechanism of ordinary core-collapse supernovae \citep[see][for a review]{kotake06}. Note that, owing to the weakness of their interaction with matter, these neutrinos carry detailed information about the dense core that cannot be obtained through photons of any frequencies. Incidentally, gravitational waves are another candidate to probe the dense core although the detection is challenging. Among various types of core-collapse phenomena, the failed supernova is as bright in neutrino emissions as ordinary core-collapse supernovae. Since the time evolutions of luminosities and spectra are qualitatively different from those of supernovae, they can be used to diagnose prompt black hole formation \citep[][]{sumi06,self09}.

The collapse of more massive stars with $M\gtrsim 260M_\odot$ has also been studied assuming that they are Population III (Pop III) stars, which were the first stars formed in the universe \citep[e.g.,][]{fryer01,self06,suwa08}. Pop III stars are thought to be very massive with $M\gtrsim 100M_\odot$ \citep[][]{nakaume01}, and they start to collapse as a result of the pair-instability during the helium-burning phase. The pair-instability is caused by the creation of electron and positron pairs, which consume some of the thermal energy to produce the rest masses of electrons and positrons. For the nonrotational case, stars with $M\gtrsim 260M_\odot$ cannot bounce against this collapse, and they form black holes directly emitting a large amount of neutrinos within a shorter ($\sim0.1$~s) duration \citep[][]{self06}. On the other hand, stars with $140M_\odot \lesssim M\lesssim 260M_\odot$ also undergo pair-instability but do not produce black holes. They reverse the collapse by rapid nuclear burning and explode into pieces; these explosions are called pair-instability supernovae \citep[e.g.,][]{woosley02}.

To investigate the formation process of black holes, the properties of hot and dense matter should be considered. In particular, meson condensation, hyperon appearance and quark deconfinement are thought to occur at supranuclear density, affecting the equation of state (EOS) markedly. The effects of possible phase transitions were once studied for core-collapse supernovae \citep[][]{mtaka85,gent93}. The evolutions of proto--neutron stars including exotic matter have been already studied by \citet{pons01a} to investigate kaon condensation, by \citet{keil95} to investigate hyperon appearance and by \citet{pons01b} to investigate quark deconfinement. On the other hand, prompt black hole formation by failed supernovae has been recently studied by \citet{sumi09} utilizing the EOS including hyperons \citep[][]{ishi08}. An EOS including quarks and pions was constructed and applied to the collapse of Pop III stars with $100M_\odot$ in our previous study \citep[][]{self08a}. While it was concluded that quarks and pions accelerate the collapse and shorten the duration of neutrino emission, it remains to be seen whether these characteristics hold for a wide range of progenitor masses. The purpose of this study is to assess the impact of quarks and pions in black hole formation treating general relativity and neutrinos for more general cases as the first serious attempt.

In this paper, we investigate the gravitational collapse and neutrino emission of black hole progenitors in a spherically symmetric model taking into account the EOS including quarks and pions proposed by \citet{self08a}. Since, as mentioned above, black hole progenitors have a broad mass range, we also study the dependence on initial mass. Moreover, observational aspects of emitted neutrinos are addressed following our other previous study \citep[][]{self08b}. This paper is organized as follows. In Section~\ref{setup}, we briefly describe the EOS's, initial models and numerical methods. The main results are reported in Section~\ref{result}. We give the numerical results for the dynamics and neutrino signal of the reference models with $40M_\odot$ in Section~\ref{refmod}, for the investigation of initial mass dependence in Section~\ref{moddep}, and for issues related to neutrino observation in Section~\ref{neudet}. Finally, Section~\ref{concl} is devoted to a summary and discussion.

\section{Setup} \label{setup}

\subsection{Equation of State} \label{eos}

In this study, we adopt the EOS formulated by \citet{self08a}, which includes the hadron--quark phase transition for finite temperatures. For the hadronic phase, this EOS utilizes a table constructed by \citet{shen98a,shen98b} based on relativistic mean field theory with the effects of thermal pions added to their table. The MIT bag model of the deconfined three-flavor strange quark matter \citep[][]{bag74} is used for the quark phase, and the hadron--quark mixed phase is obtained from the Gibbs conditions in the EOS. In this case, the substance is composed not only of $u$ quarks but also of $d$ quarks, and there is an essential difference from the phase transition of a single substance such as the liquid-vapor transition of H$_2$O \citep[e.g.,][]{glende92}. For instance, the pressure of the hadron--quark mixed phase is not constant in an isothermal process, as shown in Figure~\ref{eos}. In the following, we refer to the EOS without pions and quarks (the original Shen EOS), the EOS without pions and with quarks, the EOS with pions and without quarks, and the EOS with pions and quarks as OO, OQ, PO, and PQ, respectively.

In our hadronic EOS, thermal pions are treated in the minimum model which assumes that their effective mass is equal to their rest mass in vacuum. In reality, pions at rest ($\mathbf{p}=0$) are subjected to a repulsive potential in the nucleons and their effective mass becomes larger than that in vacuum. In this case, the pion population is suppressed. Thus, our model corresponds to an extreme case where pions are overproduced, provided that the $p$-wave $\pi N$ attraction is omitted \citep[][]{onsh09}. Recently, it has also been pointed out that hyperons play an important role in black hole formation by stellar collapse \citep[][]{ishi08,sumi09}. Hyperons are not included in our hadronic EOS; however, we are planning to investigate their effects in a future work. Incidentally, nuclei except for $\alpha$-particles are treated as a single species in the Thomas--Fermi approximation. It is preferable to adopt the EOS in nuclear statistical equilibrium (NSE) or, more sophisticatedly, in non-NSE abundances determined from the preceding quasistatic evolutions, particularly  for the temperature $T \lesssim 0.5$~MeV \cite[][]{fischer09}.

In the MIT bag model, free quarks are confined in the ``bag'', and this bag has a positive potential energy per unit volume, $B$. This parameter is called the bag constant and characterizes the model. For instance, the larger the value of $B$, the higher the transition density and temperature. \citet{self08a} have shown that the EOS with $B\gtrsim250$~MeV~fm$^{-3}$ (in another unit, $B^{1/4}\gtrsim209$~MeV) is consistent with recent observations of compact stars. In the present study, we set $B=250$~MeV~fm$^{-3}$. Note that the end point of the transition line (the so-called critical point) has been suggested to reside in the temperature range 150~MeV~$\le T_c \le 200$~MeV on the basis of the heavy-ion collision experiments and lattice QCD calculations. If this is the case, the quark matter should be most stable for all densities at a temperature of $> \!\! 200$~MeV and the hadron--quark transition should occur at $< \!\! 150$~MeV. Our model is consistent with this picture, although it cannot reproduce the critical point in principle. We can confirm the validity of our model from the phase diagrams and free energies, which are shown in Figures 5 and 6 of \citet{self08a}, respectively. Very recently, dealing with quark matter in the manner described above, \citet{sagert09} performed simulations of the successful supernova explosions of 10 and $15M_\odot$ progenitors using spherical models. An important difference in their EOS from ours is that their bag constant was very small, $B=90$~MeV~fm$^{-3}$ ($B^{1/4}=162$~MeV). Further studies are necessary to fix the bag constant.

\subsection{Initial Models} \label{init}

In this study, we investigate the collapse of stars with various masses. The adopted models are those of a $40M_\odot$ star with solar metallicity proposed by \citet{woosley95}, a Pop III star with $100M_\odot$ proposed by \citet{nomoto05} and a Pop III star with $375M_\odot$ proposed by \citet{self06}. The former two models were constructed from evolutionary calculations, while the latter model was obtained from the equilibrium configuration of a gravitationally unstable iron core. The evolution scenario of the model with $375M_\odot$ is different from that of the other two models. The $40M_\odot$ star with solar metallicity and the $100M_\odot$ Pop III star undergo successive nuclear burning and form an iron core at the end of quasistatic evolution. This iron core becomes gravitationally unstable owing to the photodisintegration of iron and starts to collapse. These features are similar to those in the case of ordinary supernova progenitors. On the other hand, the Pop III star with $375M_\odot$ starts to collapse through pair-instability, and an iron core is formed during the collapse. Therefore, an ad hoc model is adopted in this study; however, its validity has been verified by \citet{self06} on the basis of the results of recent evolutionary calculations \citep[][]{fryer01,ohkubo06}. 

Note that the collapses of all models adopted here have already been examined for the EOS model~OO, namely, the EOS proposed by \citet{shen98a,shen98b}, using the same numerical methods as those utilized in this paper. The models of stars with 40, 100 and $375M_\odot$ were studied by \citet{sumi07}, \citet{self07}, and \citet{self06}, respectively. Moreover, results for the collapse of the $100M_\odot$ star under EOS models~PO, OQ and PQ are given in \citet{self08a}. The collapses of these three progenitor models result in black hole formation under EOS model OO, although their dynamical features are different from each other. The $40M_\odot$ star produces a bounce before black hole formation owing to the nuclear force because the central density exceeds the nuclear density. The $100M_\odot$ star also produces a bounce but owing to the thermal pressure of nucleons. The entropy of this model is sufficiently high for nuclei to dissociate into nucleons and $\alpha$-particles at subnuclear density. Then the thermal pressure of nucleons and $\alpha$-particles produces a weak bounce. On the other hand, the $375M_\odot$ star does not produce a bounce and collapses to a black hole directly. In this study, we examine whether or not these features are qualitatively and/or quantitatively changed by the effects of quarks and pions through comparisons with the results of previous studies \citep[][]{self06,self07,sumi07}.

\subsection{Numerical Methods} \label{code}

The general relativistic implicit Lagrangian hydrodynamics code, which simultaneously solves the neutrino Boltzmann equations \citep[][]{yamada97,yamada99,sumi05}, is utilized to compute the dynamics of spherical gravitational collapse and neutrino transport. This code can solve the evolution of space time as well as the dynamics up to black hole formation. Since the event horizon has been proved to always be located outside the apparent horizon, black hole formation can be confirmed by finding the apparent horizon. For the Misner--Sharp metric \citep[][]{misner64}, which is spherically symmetric and adopted in our code, the radius of the apparent horizon is simply written as $r=2G\widetilde m/c^2$, where $c$ and $G$ are the velocity of light and the gravitational constant, respectively \citep[][]{riper79}. $r$ and $\widetilde m$ are the circumference radius and the gravitational mass, respectively, and we solve them as functions of time and the baryon mass coordinate. Therefore, the appearance of the apparent horizon can be concluded when the coordinate at which $r = 2G\widetilde m/c^2$ is satisfied appears in our numerical simulations \citep[see][for details]{yamada97,self06}.

To obtain the neutrino distribution functions, we solve the Boltzmann equations by a finite difference scheme ($S_N$ method) on discretized grid points for the radial Lagrangian coordinate, neutrino energy spectrum, and neutrino angular distribution. In our simulation, we consider four species of neutrino, $\nu_e$, $\bar\nu_e$, $\nu_\mu$ and $\bar\nu_\mu$, assuming that the distribution function of $\nu_\tau$ ($\bar\nu_\tau$) is equal to that of $\nu_\mu$ ($\bar\nu_\mu$). For the collision terms of the Boltzmann equations, we calculate the scattering kernels explicitly in terms of the angles and energies of incoming and outgoing neutrinos \citep[see][for details]{mezza93,yamada99}. The neutrino reactions taken into account are (1) electron-type neutrino absorption on neutrons and its inverse, (2) electron-type antineutrino absorption on protons and its inverse, (3) neutrino scattering on nucleons, (4) neutrino scattering on electrons, (5) electron-type neutrino absorption on nuclei, (6) neutrino coherent scattering on nuclei, (7) electron--positron pair annihilation and creation, (8) plasmon decay and creation and (9) neutrino bremsstrahlung. We adopt the reaction rate for (8) from \citet{braaten93}, that for (9) from \citet{maxwell87} and that for the others reactions from \citet{bruenn85}.

Note that the neutrino treatments described above must be changed for the region where the phase transition occurs. This is because our EOS is constructed assuming that electron-type neutrinos are in equilibrium with other particles in the hadron--quark mixed phase and the pure quark phase \citep[][]{self08a}. Under this assumption, we set the neutrino distribution functions to be Fermi--Dirac functions for all species, and the electron-type lepton fraction, $Y_l$, is assumed to be conserved for each fluid element, instead of solving the transport and reaction of neutrinos to compute the time evolutions of neutrino distribution functions and the electron fraction, $Y_e$. Moreover, we neglect the entropy variation resulting from the neutrino transport. These modifications are only applied at the mesh points where quarks appear. We can justify this simplification in the neutrino treatments because the density is sufficiently high for neutrinos to be trapped at the phase transition. In fact, as shown later, neutrinos trapped inside the quark core cannot escape because a black hole is formed suddenly $< \!\! 1$~ms after the phase transition. Note that $\beta$ equilibrium is also used to determine the fraction of $s$-quarks because strangeness is generated only by weak interactions such as $s \leftrightarrow u + e^- + \bar \nu_e$. If $\beta$ equilibrium is not assumed, the three-flavor quark EOS becomes a function not only of density, temperature and $Y_e$ (or $Y_l$) but also of strangeness.

In this study, the numbers of mesh points for the radial Lagrangian coordinate, neutrino energy spectrum and neutrino angular distribution are chosen to coincide with those of previous studies given in Section~\ref{init}. For instance, for the $40M_\odot$ model, we use 255 mesh points for the radial Lagrangian coordinate and 14 and 6 mesh points for the energy spectrum and angular distribution, respectively, which are the same as those in \citet{sumi07}. The uncertainties originating from the resolutions were evaluated to be $\sim \!\! 10$\% by \citet{self07}. Note that rezoning and dezoning of the Lagrangian coordinate are performed during the simulations \citep[][]{sumi05}. Since sufficiently high resolution is needed in the vicinity of the shock wave, rezoning is performed for the accreting regions. On the other hand, dezoning is performed for the inner regions of proto--neutron stars where materials are almost hydrostatic.

\section{Results} \label{result}

\subsection{Collapse of $40M_\odot$ Star} \label{refmod}

In this section, we examine the collapse of $40M_\odot$ star as the reference models. In the previous study \citep[][]{sumi07}, the model with EOS~OO was shown to produce a bounce owing to the nuclear force before black hole (apparent horizon) formation. Neutrinos are emitted mainly during the period from the bounce to black hole formation. In this study, we find that these qualitative features are not changed for the models including quarks and/or pions. However, quantitative differences appear in, for instance, the time interval between the bounce and black hole formation. In the following, we investigate this phase in detail.

We show the time profiles of the central baryon mass density in Figure~\ref{cd40}. The bounce owing to the nuclear force corresponds to the spikes at $t=0$, which is defined as the time of the bounce. At this moment, EOS dependence does not appear because the hadron--quark phase transition has not yet occurred and the contribution of pions is still minor. Although the bounce produces a shock wave, it does not propagate out of the core and is stalled. At the center, a proto--neutron star is formed and gradually contracts owing to the accretion of shocked matter. This phase corresponds to the gradual density increase in Figure~\ref{cd40}. Finally, the core collapses to a black hole and the central density increases rapidly. We can see that the time interval from the bounce to black hole formation is reduced owing to the contribution of quarks and pions. This is because the EOS becomes softer and the maximum mass for the stable configurations of proto--neutron stars decreases as found in earlier studies. In fact, the time intervals are 1.049~s, 1.086~s, 1.145~s, and 1.345~s for the models with EOS's~PQ, OQ, PO, and OO, while the maximum masses of the ``cold'' neutron stars are $1.8M_\odot$, $1.8M_\odot$, $2.0M_\odot$, and $2.2M_\odot$, respectively \citep[][]{self08a}. Since the mass accretion rate does not differ among the EOS models, a soft EOS leads to a reduction of the time interval. Note that we cannot simply quote the maximum masses of cold neutron stars because the proto--neutron star is hot and lepton rich. However, our results are roughly consistent with this trend.

While both quarks and pions soften the EOS and promote black hole formation, their effects are qualitatively different. Comparing the models with EOS's~PO and PQ (or EOS's~OO and OQ), we can see that quarks have an effect in the very late phase. In other words, the transition to the mixed phase triggers the collapse to a black hole. The mass--radius relations of a neutron star using our EOS's are shown in Figure~7 of \citet{self08a}. From this figure, the maximum mass for the mixed EOS is very close to the mass at which the hadron--quark phase transition makes a difference. This description is consistent with the fact that black hole formation occurs immediately after the appearance of quarks. On the other hand, the effect of pions begins to appear as a gradual increase in the density because the thermal pions appear before pion condensation.

The features of the appearance of quarks and pions can be seen in Figure~\ref{yi40}, where the profiles of the particle fractions and the baryon mass density of the model with EOS~PQ are shown for each step. Note that there are nuclei and $\alpha$-particles in the outer region whose profiles are not shown in Figure~\ref{yi40}. When the central density is $4\times10^{14}$~g~cm$^{-3}$ ($t=272$~ms, or 777~ms before black hole formation: upper left panel of Figure~\ref{yi40}), the population of thermal pions is small and they do not affect the dynamics. We can also confirm this from the comparison of EOS~OQ and EOS~PQ in Figure~\ref{cd40}. When the central density is $8\times10^{14}$~g~cm$^{-3}$ ($t=1022$~ms, or 27~ms before black hole formation: upper right panel of Figure~\ref{yi40}), pion condensation has already occurred in the central region; however, quarks have not appeared yet. When the central density is $2.5\times10^{15}$~g~cm$^{-3}$ (0.07~ms before black hole formation: lower left panel of Figure~\ref{yi40}), quarks in the mixed phase prevail in the central region. In this phase, the star is already collapsing dynamically to a black hole, which is consistent with the fact that quarks begin to have an effect in the very late phase. At the time of black hole formation (lower right panel of Figure~\ref{yi40}), the central density increases to $1.5\times10^{16}$~g~cm$^{-3}$ and the pure quark matter resides in the central region.

We show the time evolutions of the density, temperature, entropy per baryon, electron fraction and radial velocity profiles of the model with EOS~PQ in Figure~\ref{ev40}, where the initial location of the apparent horizon and the profiles of sound speed with opposite sign at the time of black hole formation are also shown. The density rises rapidly after the phase transition. In contrast, the temperature profile at the moment of black hole formation has a multi-peaked shape. This will be discussed again later. As can be recognized from the entropy profiles, the collapse to a black hole is adiabatic. The electron fraction decreases during the phase transition owing to the generation of $s$-quarks, which have negative charge. In the pure quark phase, matter is compressed without changing the fractions of not only electrons but also other particles owing to the chemical equilibrium. Note that, in this regime, quarks and leptons can be regarded as degenerate and relativistic ideal Fermi gases, where the number density $n_i$ and chemical potential $\mu_i$ relate as $n_i \propto \mu_i^3$. Comparing the radial velocity and sound speed at the time of black hole formation, we can see that the infall becomes subsonic for the pure quark phase, while it is supersonic for the hadron--quark mixed phase (see also lower right panel of Figure~\ref{yi40}). The sound speed is lower for the mixed phase, which corresponds to the gradual increase in the pressure against contraction in Figure~\ref{eos}. Therefore, the infall velocity becomes maximum in the region with the mixed phase and does not increase further in the region with the pure quark phase. Note that pure quark matter in the MIT bag model is asymptotically close to a relativistic ideal gas at the high-density limit. Thus, the sound speed is nearly $c/\sqrt{3}$ for the innermost region.

The evolution of the central density and temperature of the model with EOS~PQ is plotted with the phase diagram in Figure~\ref{dent-ph}. To draw the phase diagram, the electron-type lepton fraction is fixed to $Y_l=0.3$, which is the same as the value at the center of this model. From this figure, we can see that the temperature decreases in the mixed phase despite the increase in density. Although this appears unfamiliar, we can interpret it in the context of a phase transition \citep[][]{mueller97}. Here we assumed that the transition is first order, although some authors have regarded it as a second-order or crossover transition \citep[e.g.,][]{aoki06}. In the first-order transition, the release of latent heat occurs. Thus, in an isothermal phase transition, the entropy varies. Note that, in our EOS, the low-density (hadron) phase has lower entropy, which is opposite to an ordinary liquid--vapor transition (e.g., water vapor has a lower density and higher entropy than liquid water in the liquid-vapor transition of H$_2$O). This means that the entropy increases during the isothermal transition. However, in our case, the entropy does not vary because matter in the collapsing star is compressed adiabatically. If the onset of the transition point is fixed, the entropy is larger for the isothermal transition than for the adiabatic transition (see Figure~\ref{schem}). Therefore, the temperature is lower for the adiabatic transition than for the isothermal transition. This is the reason for the temperature decrease during the phase transition.

The multi-peaked shape of the temperature profile in Figure~\ref{ev40} mentioned earlier is due to the temperature decrease during the phase transition. For instance, the boundary between the hadronic phase and mixed phase is at approximately $\sim \!\! 1.4 M_\odot$ at the moment of black hole formation (lower right panel of Figure~\ref{yi40}). Since the temperature of each Lagrangian fluid element (mass coordinate) decreases for the mixed phase, one peak is generated near the boundary (see the solid line in Figure~\ref{ev40}). On the other hand, a local minimum appears near the boundary between the mixed phase and quark phase ($\sim \!\! 1.1 M_\odot$) because the temperature increases again in the pure quark phase.

We now turn to neutrino emission. In Figure~\ref{neuprof}, the average energies and luminosities of neutrinos are shown as a function of time. Note that $\nu_\mu$ and $\bar\nu_\mu$ have the same type of reactions, the difference in coupling constants is minor and, as already mentioned, $\nu_\tau$ ($\bar\nu_\tau$) is assumed to be the same as $\nu_\mu$ ($\bar\nu_\mu$). Therefore, we collectively denote these four species as $\nu_x$. The average energy presented here is defined by the rms value. Comparing the models with EOS's~OO and OQ or the models with EOS's~PO and PQ, we can see that the profiles do not significantly differ from each other up to the time of black hole formation for the models with quarks. This is because, again, quarks only have an effect at the final moment. As a result, the total energies of emitted neutrinos for the models including quarks are lower than those of the models without quarks because of the shorter durations of neutrino emission.

A similar trend can be seen in the comparison between the models with and without pions. However, pions increase the average energies and luminosities of neutrinos gradually. Roughly speaking, neutrinos can be regarded as being emitted from the neutrino sphere, where the optical depth is $2/3$ for neutrinos with a typical energy. The neutrino luminosity summed over all species is roughly given by the accretion luminosity $L^\mathrm{acc}_\nu \sim GM_\nu \dot{M} / R_\nu$ \citep[][]{thomp03}, where $R_\nu$, $\dot{M}$, and $M_\nu$ are the radius of the neutrino sphere, the mass accretion rate and the mass enclosed by $R_\nu$, respectively. The average energy of neutrinos is approximately proportional to the temperature of the neutrino sphere, $T_\nu$. Owing to appearance of pions, the EOS becomes soft and the inner core contracts as shown in Figure~\ref{cd40}. Therefore, $R_\nu$ decreases and $T_\nu$ increases. This is why the average energies and luminosities of neutrinos are increased by pions. This effect is particularly notable for $\nu_x$, because $\nu_x$ does not have charged-current reactions. The absence of charged-current reactions makes the core optically thinner and the radius of the neutrino sphere smaller for $\nu_x$. Thus, the signal of $\nu_x$ is more sensitive to the difference in the inner region, that is, the appearance of pions.

\subsection{Mass Dependences} \label{moddep}

In this section, we show the results for the Pop III stars with $100M_\odot$ and $375M_\odot$ and compare them with those for the $40M_\odot$ star with the solar metallicity given in Section~\ref{refmod}. As mentioned already, results for the models with $100M_\odot$ have also been reported by \citet{self08a}. We show the time profiles of the central baryon mass density for the models with $100M_\odot$ and $375M_\odot$ in Figure~\ref{cdmd}. Note that the time is measured from the point when the central density exceeds $10^{12}$~g~cm$^{-3}$ for the models with $375M_\odot$ because the star collapses to a black hole directly without a bounce. For the models with $100M_\odot$, the time is measured from the bounce, similarly to the models with $40M_\odot$.

The EOS dependence of the models with $100M_\odot$ is similar to that for the models with $40M_\odot$, although there are some differences. First, the central density at the bounce is $\sim \!\! 2\times10^{14}$~g~cm$^{-3}$ for the $100M_\odot$ models, while it is $\sim \!\! 3.2\times10^{14}$~g~cm$^{-3}$ for the $40M_\odot$ models. This is because the bounce mechanism is different. As already stated, the $100M_\odot$ models produce a bounce owing not to the nuclear force but to the thermal pressure of nucleons. Second, the effect of pions is clearer for the $100M_\odot$ models for the following reason. The entropy in the central region of a $100M_\odot$ star ($\sim \!\! 3.5 k_\mathrm{B}$ per baryon) is higher than that of a $40M_\odot$ star ($\sim \!\! 1 k_\mathrm{B}$ per baryon), where $k_\mathrm{B}$ is the Boltzmann constant, and the temperature is also higher. In Figure~\ref{dent-ph100}, we plot the evolution of the central density and temperature for the $100M_\odot$ model with EOS~PQ with the phase diagram for $Y_l=0.17$, which is the value at the center of this model. Upon comparison with Figure~\ref{dent-ph}, one can recognize that a $100M_\odot$ star has a higher temperature than a $40M_\odot$ star. Therefore, a $100M_\odot$ star has a larger population of pions than a $40M_\odot$ star. Aside from this difference, the fact that quarks and pions hasten black hole formation and reduce neutrino emission is unchanged for the models with $100M_\odot$.

For the models with $375M_\odot$, the EOS dependence is not strong as shown in Figure~\ref{cdmd}. In this case, the core does not bounce and is already collapsing to a black hole at the time of quark and/or pion appearance. Therefore, quarks and pions have a limited effect. In Figure~\ref{dv375}, where profiles of the density and radial velocity at the time of apparent horizon formation are shown with the locations of the apparent horizon, we can see that quarks accelerate the collapse and affect the innermost region inside the apparent horizon. Since, as already mentioned, the apparent horizon is always located inside the event horizon, quarks do not affect the dynamics and neutrino signals outside the event horizon (black hole). Thus, the results and discussion for the neutrino emissions of Pop III stars reported in \citet{self06} do not require modification.

\subsection{Implications for Neutrino Detection} \label{neudet}

When quarks and pions have an impact on black hole formation, the neutrino signals detected by terrestrial neutrino detectors will be affected. Among the initial models investigated in this study, a difference in the neutrino signal may occur for $40M_\odot$ and $100M_\odot$ stars. Our $100M_\odot$ star is that of a Pop III star, which is a first-generation star and no longer exists survive in the nearby universe. Therefore, in this section, we investigate the detectability of neutrinos emitted during black hole formation by our $40M_\odot$ models for the currently operating neutrino detector SuperKamiokande. In the evaluation of neutrino event numbers, we take into account neutrino oscillation following \citet{self08b}.

Neutrinos emitted from the stellar core-collapse propagate through the stellar envelope, where neutrino flavor conversion occurs by the Mikheyev--Smirnov--Wolfenstein effect. When neutrinos pass through the earth before detection, they also undergo flavor conversion inside the earth. In this case, the results of flavor conversion depend on the nadir angle of the progenitor. In our analyses, we utilize realistic profiles of the progenitor \citep[][]{woosley95} and the earth \citep[][]{dziewo81}. There are two undetermined parameters of neutrino mixing, namely, the mixing angle $\theta_{13}$ and the mass hierarchy. For the former, only the upper limit is given as $\sin^2\theta_{13}\leq2.0\times10^{-2}$, while the other mixing angles have been well measured. Whether the sign of the mass-squared difference $\Delta m^2_{31}$ is plus (normal mass hierarchy) or minus (inverted mass hierarchy) is also currently uncertain. In this study, we investigate the dependence of the neutrino event number on these undetermined mixing parameters as well as the nadir angle of the progenitor. The distance from the progenitor, $R$, also determines the event number, which can be scaled simply as $\propto 1/R^2$. Here we set $R=10$~kpc, which is a typical length for our Galaxy.

In Figure~\ref{ndettot}, the time-integrated event numbers of the collapse of the $40M_\odot$ models are shown for various parameter sets as well as for the EOS models. We can see that the expected event number is sufficiently large for all cases, being similar to that of ordinary supernova neutrinos, $\sim \!\! 10,000$ \citep[e.g.,][]{taka01}. The error bars in Figure~\ref{ndettot} represent the upper and lower limits owing to different nadir angles. In the case with the earth effects, the energy spectral shape is deformed to a wavelike shape \citep[see Figure~7 of][for instance]{self08b} because the typical length of neutrino oscillation becomes comparable to the size of the earth and the neutrino survival probability becomes sensitive to the neutrino energy. However, integrating over the neutrino energy, the fluctuation is smoothed out and its impact on the total event number is small. For all EOS models, the event number decreases for the inverted mass hierarchy with larger $\sin^2\theta_{13}$ because almost all $\bar\nu_e$, which has the greatest contributions to the event number through the inverse beta decay reaction, converts to $\bar\nu_\mu$ or $\bar\nu_\tau$ for this parameter set.

From Figure~\ref{ndettot}, we can see that the emergence of quarks and pions affects the total neutrino event number in addition to the duration of neutrino emission for some cases. The neutrino event number of the model with EOS~OQ is $\gtrsim \!\! 30$\% smaller than that of the model with EOS~OO for both parameter sets of neutrino oscillation. This difference arises mainly from the short duration of the neutrino emission. In fact, as shown in Figure~\ref{neuprof}, the duration for the model with EOS~OQ is $\lesssim \!\! 20$\% shorter than that for the model with EOS~OO, where the luminosity and average energy increase in the late phase. While quarks merely affect the duration, pions increase the luminosity and average energy of neutrinos before black hole formation. Therefore, the neutrino event number of the model with EOS~PO is not greatly reduced from that of the model with EOS~OO. Moreover, the model with EOS~PQ is almost identical with the model with EOS~OQ. This similarity can be clearly seen in Figure~\ref{ndetspc}, which shows the energy spectra of the time-integrated event number. Since the model with EOS~PQ has a shorter duration but a higher luminosity and average energy of neutrinos than the model with EOS~OQ, both effects cancel out.

\section{Summary and Discussion} \label{concl}

In this study, we have performed a series of black hole-forming core-collapse simulations involving EOS's with quarks and/or pions. Our EOS's utilize the MIT bag model with $B=250$~MeV~fm$^{-3}$ for the quark phase. We performed numerical computations by solving the Boltzmann equation with hydrodynamics under spherical symmetry in general relativity to obtain detailed information on the energy spectrum of neutrino emission during the evolution. We adopted initial models with three different masses, namely, 40, 100, and $375M_\odot$, although we mainly reported results for a $40M_\odot$ model. We found that quarks and pions shorten the duration of neutrino emission if the collapse bounces before black hole formation. In particular, pions also increase the luminosity and average energy of neutrinos before black hole formation. For the Galactic events of $\sim \!\! 40M_\odot$ stars, the neutrino events at the currently operating detector SuperKamiokande are not only detectable but also affected by the emergence of quarks and pions. The observational features of the model with quarks and pions are almost the same as those of the model with quarks but without pions. Note that, the hadron--quark phase transition gives rise to the nonmonotonic increase in the temperature during the collapse.

Needless to say, there are some issues beyond the scope of this study. Among other issues, there are many other possible EOS's of hot and dense matter. In particular, the duration of neutrino emission becomes short and the luminosity and average energy of neutrinos increase also for soft hadronic EOS's. To probe the emergence of quarks and pions for the Galactic events, more detailed statistical analyses will be needed \citep[][]{self09}. Hyperons are thought to emerge at the densities of interest, which we have not taken into account. While core-collapse simulations involving hyperon populations have already been performed and reported in a separate paper \citep[][]{sumi09}, we are planning to construct an EOS including quarks, pions and hyperons and apply it to our numerical simulations. The consideration of mesons, not only pions but also kaons, requires sophisticated treatment. The use of a quark EOS with a different value of the bag constant is another concern. Moreover, other quark models, such as the Nambu--Jona-Lasinio model, have been proposed for the modeling of astrophysical phenomena \citep[e.g.,][]{blas05}. Note that, our result that quarks have an effect in the very late phase may change when quark models with lower transition density are adopted. Multidimensional simulations are also certainly worth investigating, although their numerical treatment of neutrino transport under general relativity is a challenging problem. Uncertainties in the evolutionary calculations of progenitor models, such as convection and massloss, will affect neutrino signals through the density profile of the outer layer \cite[][]{sumi08, fischer09}. It should be emphasized, however, that this is the first serious attempt to assess the impact of quarks and pions in the dynamical simulations of black hole formation and will serve as an important foundation for future studies.

\acknowledgments

We are grateful to Akira Ohnishi and Hideyuki Suzuki for fruitful discussions. In this work, numerical computations were partially performed on the supercomputers at Research Center for Nuclear Physics (RCNP) in Osaka University, Center for Computational Astrophysics (CfCA) in the National Astronomical Observatory of Japan (NAOJ), Yukawa Institute for Theoretical Physics (YITP) in Kyoto University, Japan Atomic Energy Agency (JAEA) and High Energy Accelerator Research Organization (KEK). This work was partially supported by Research Fellowship for Young Scientists from the Japan Society for Promotion of Science (JSPS) through 18-510 and 21-1189, and Grants-in-Aid for the Scientific Research from the Ministry of Education, Culture, Sports, Science and Technology (MEXT) in Japan through 17540267, 18540291, 18540295, 19104006, 19540252, 20105004, 21540281 and 22540296.


\clearpage

\begin{figure}
\plotone{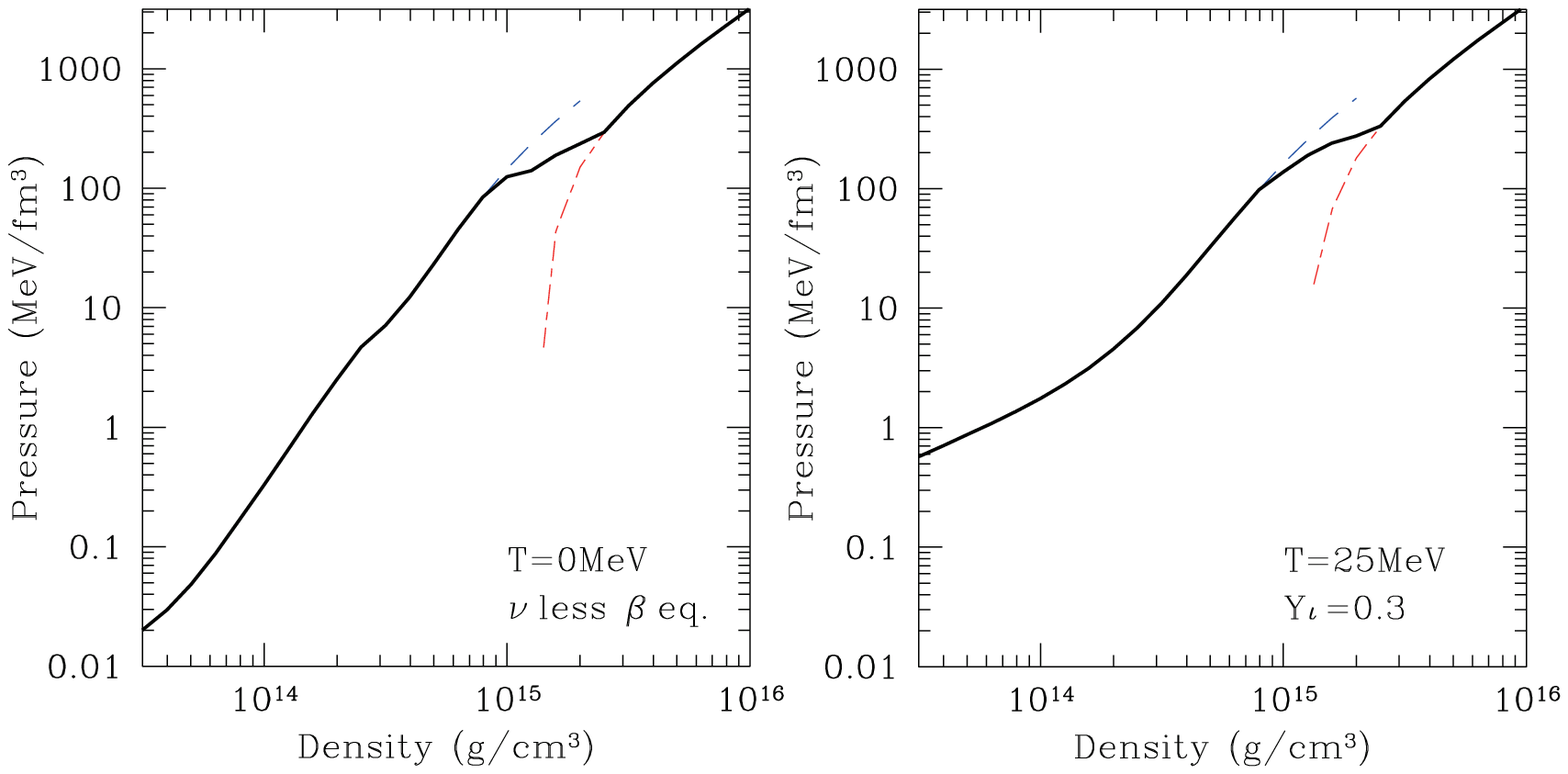}
\caption{Pressure as a function of baryon mass density for hadron-quark mixed matter (thick solid lines), pure hadronic matter (dashed lines) and pure quark matter (dot-dashed lines). The left panel corresponds to the case for matter in neutrino-less $\beta$ equilibrium at zero temperature, whereas the right panel corresponds to that with temperature $T=25$~MeV and electron-type-lepton fraction $Y_l=0.3$, where $Y_l$ is defined as the sum of the electron fraction, $Y_e$, and the electron-type-neutrino fraction, $Y_{\nu_e}$. In both the panels, the bag constant is chosen as $B=250$~MeV~fm$^{-3}$ for the quark matter and thermal pions are included in the hadronic matter.}
\label{eos}
\end{figure}

\begin{figure}
\plotone{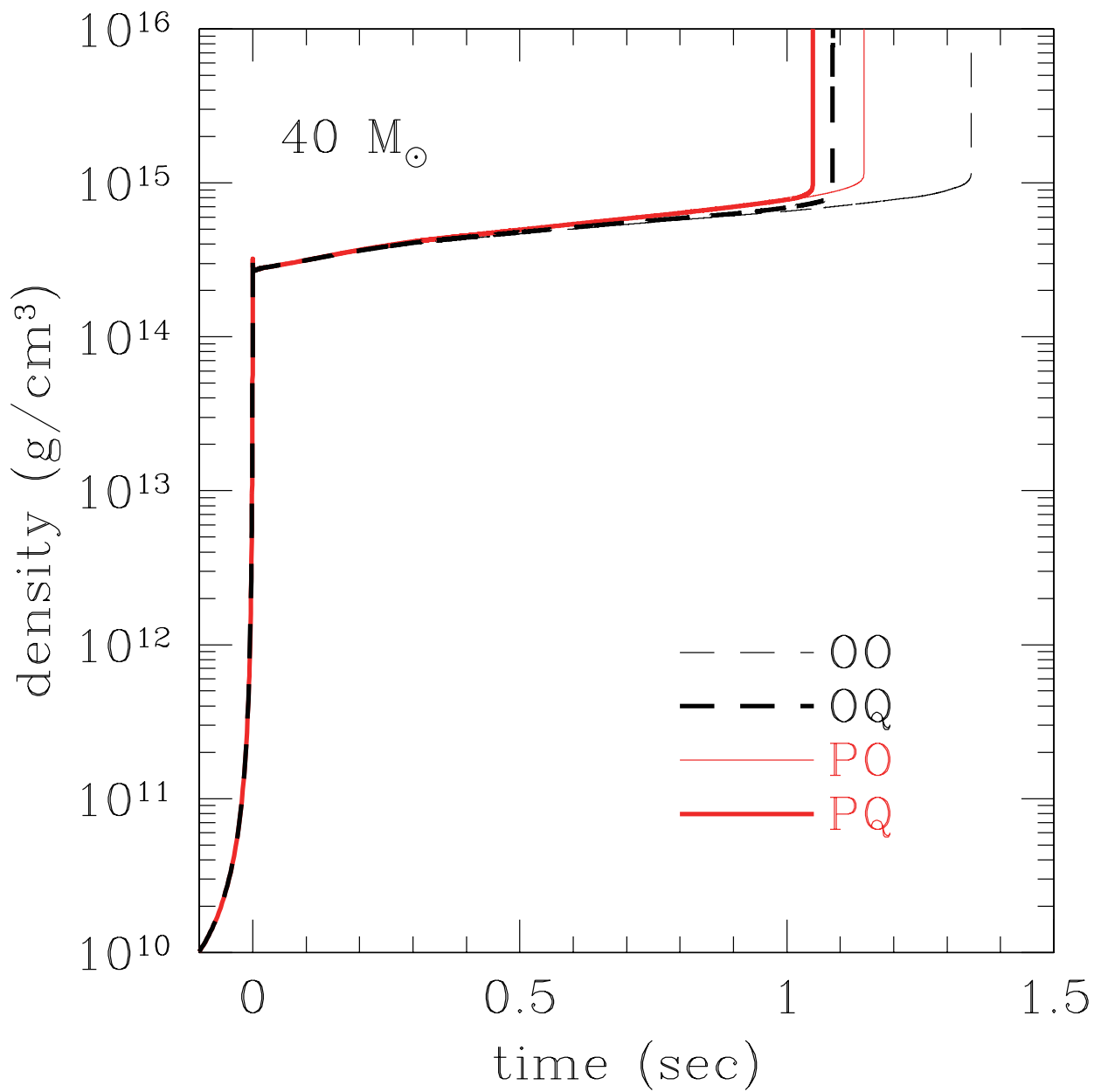}
\caption{Time profiles of the central baryon mass density for the collapse of the models with $40M_{\odot}$. Thin dashed, thick dashed, thin solid and thick solid lines correspond to the results for EOS's~OO, OQ, PO and PQ, respectively. The time is measured from the bounce. Note that the result for EOS~OO is also given in \citet{sumi07}.}
\label{cd40}
\end{figure}

\begin{figure}
\plotone{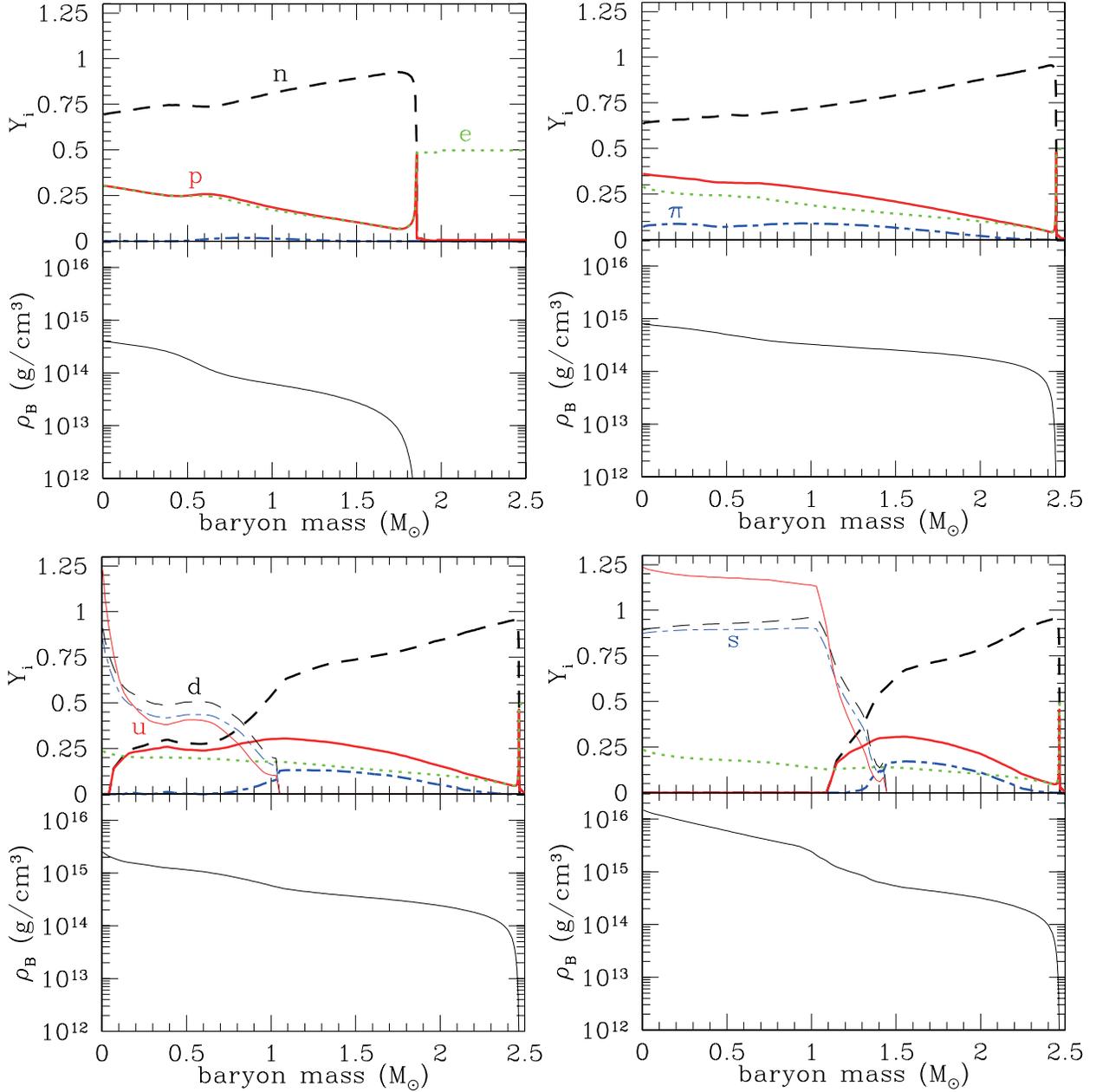}
\caption{Profiles of the particle fractions (upper plots) and baryon mass density (lower plots) of $40M_{\odot}$ model with EOS~PQ, where $Y_i \equiv \frac{n_i}{n_B}$. $n_i$ represents the number density of particle $i$, and $n_B$ represents the baryon number density. The upper left, upper right and lower left panels respectively correspond to 777~ms, 27~ms and 0.07~ms before black hole (apparent horizon) formation, whereas the lower right panel represents the moment of black hole formation. Note that 777~ms before black hole formation corresponds to 272~ms after the bounce.}
\label{yi40}
\end{figure}

\begin{figure}
\epsscale{0.6}\plotone{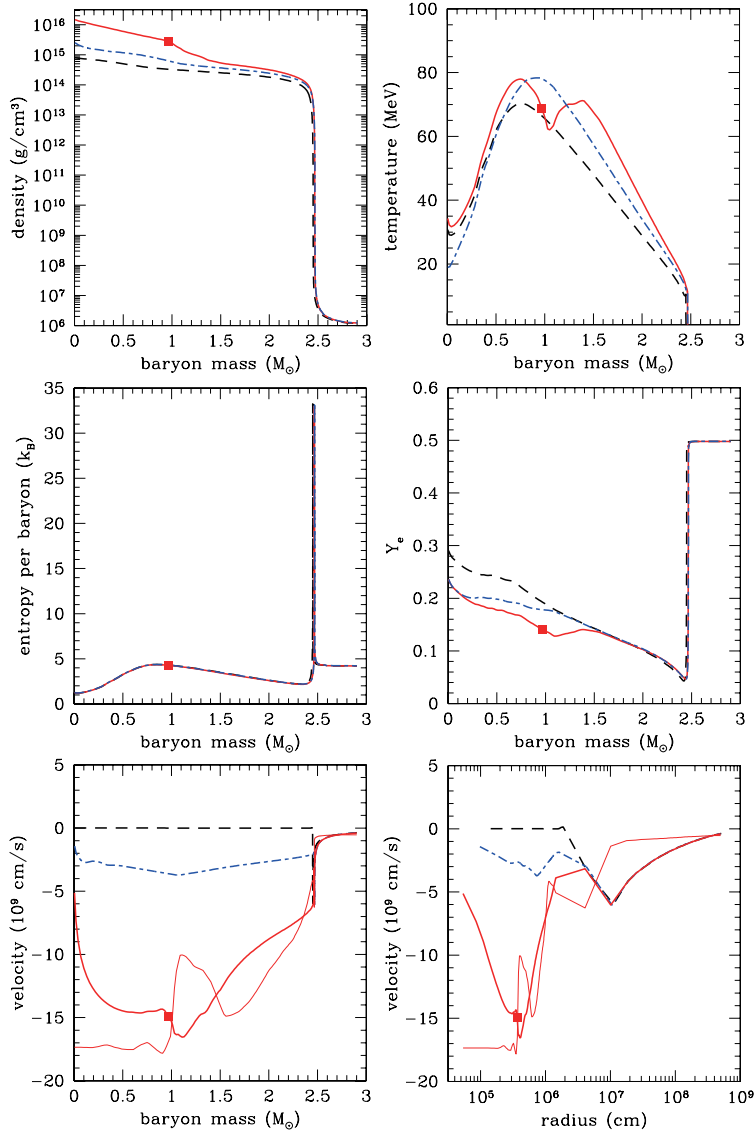}
\caption{Time evolutions of the density (upper left), temperature (upper right), entropy per baryon (middle left), electron fraction (middle right), and radial velocity (lower left and right) profiles for the $40M_{\odot}$ model with EOS~PQ. Note that plots in the lower-right panel are shown as functions of radius while plots in the other panels are functions of the baryon mass coordinate. The thick dashed and thick dot-dashed are snapshots 27~ms and 0.07~ms before black hole (apparent horizon) formation, respectively, whereas thick solid lines represent the moment of black hole formation. Squares show the initial location of the apparent horizon. In addition, profiles of the sound speed with opposite sign at the moment of black hole formation are shown as thin solid lines in the lower panels.}
\label{ev40}
\end{figure}

\begin{figure}
\epsscale{1}\plotone{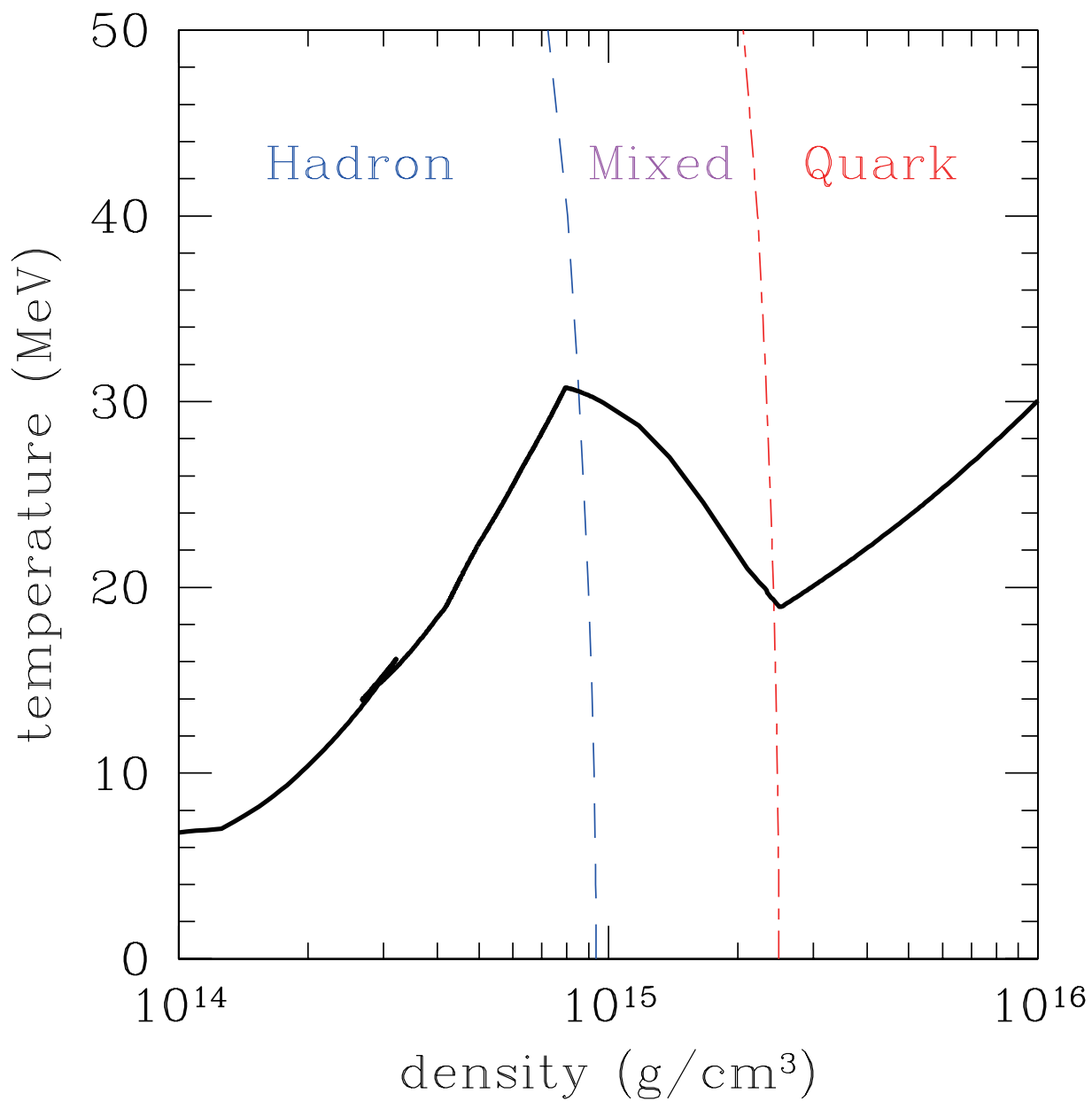}
\caption{Evolution of the central density and temperature of $40M_{\odot}$ model with EOS~PQ (thick solid line) and the phase diagram for $Y_l=0.3$. The dashed line represents the boundary between hadronic matter and mixed matter, and the dot-dashed line represents that between mixed matter and quark matter. The point of inflection at $\sim \!\! 3\times10^{14}$~g~cm$^{-3}$ is due to the bounce.}
\label{dent-ph}
\end{figure}

\begin{figure}
\plotone{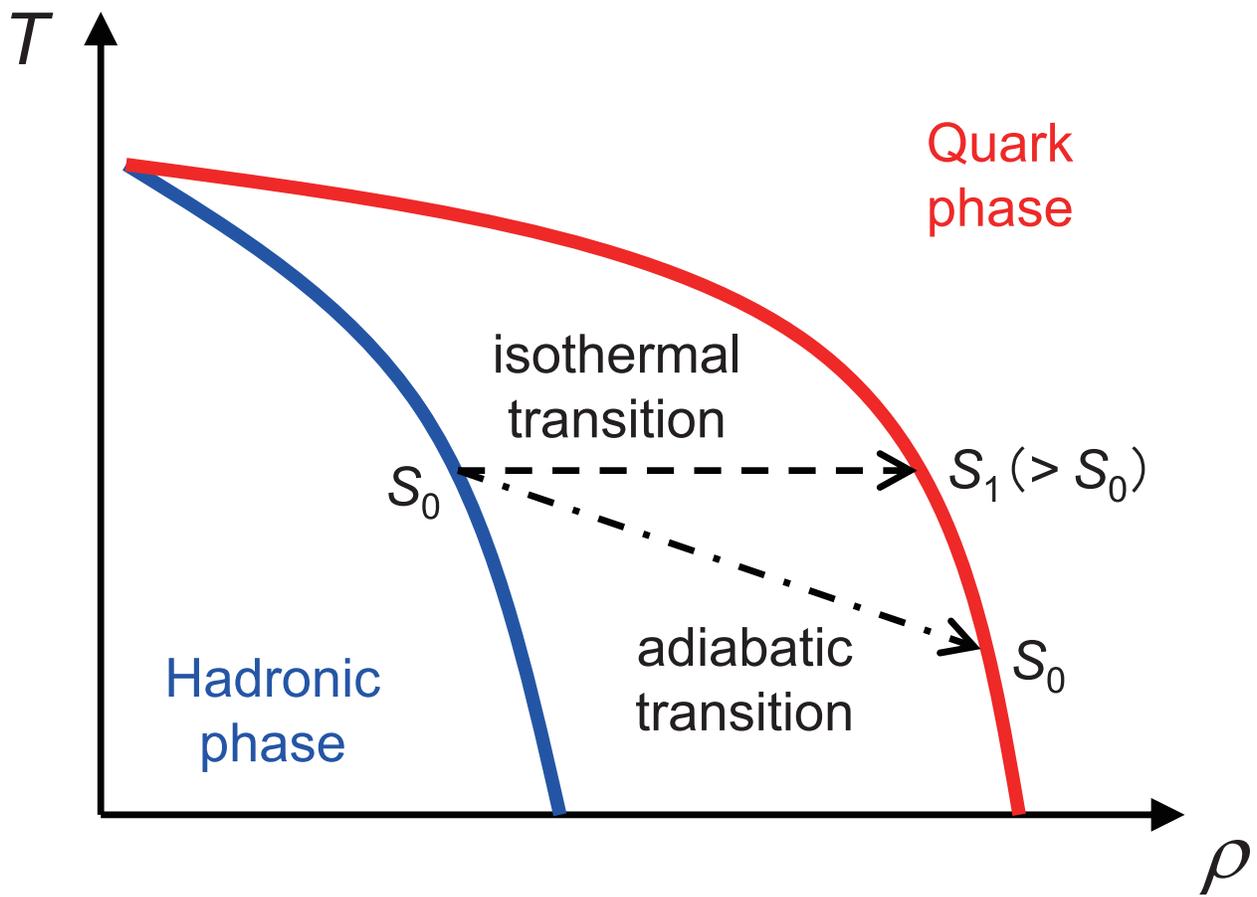}
\caption{Schematic picture of the phase transition.}
\label{schem}
\end{figure}

\begin{figure}
\plotone{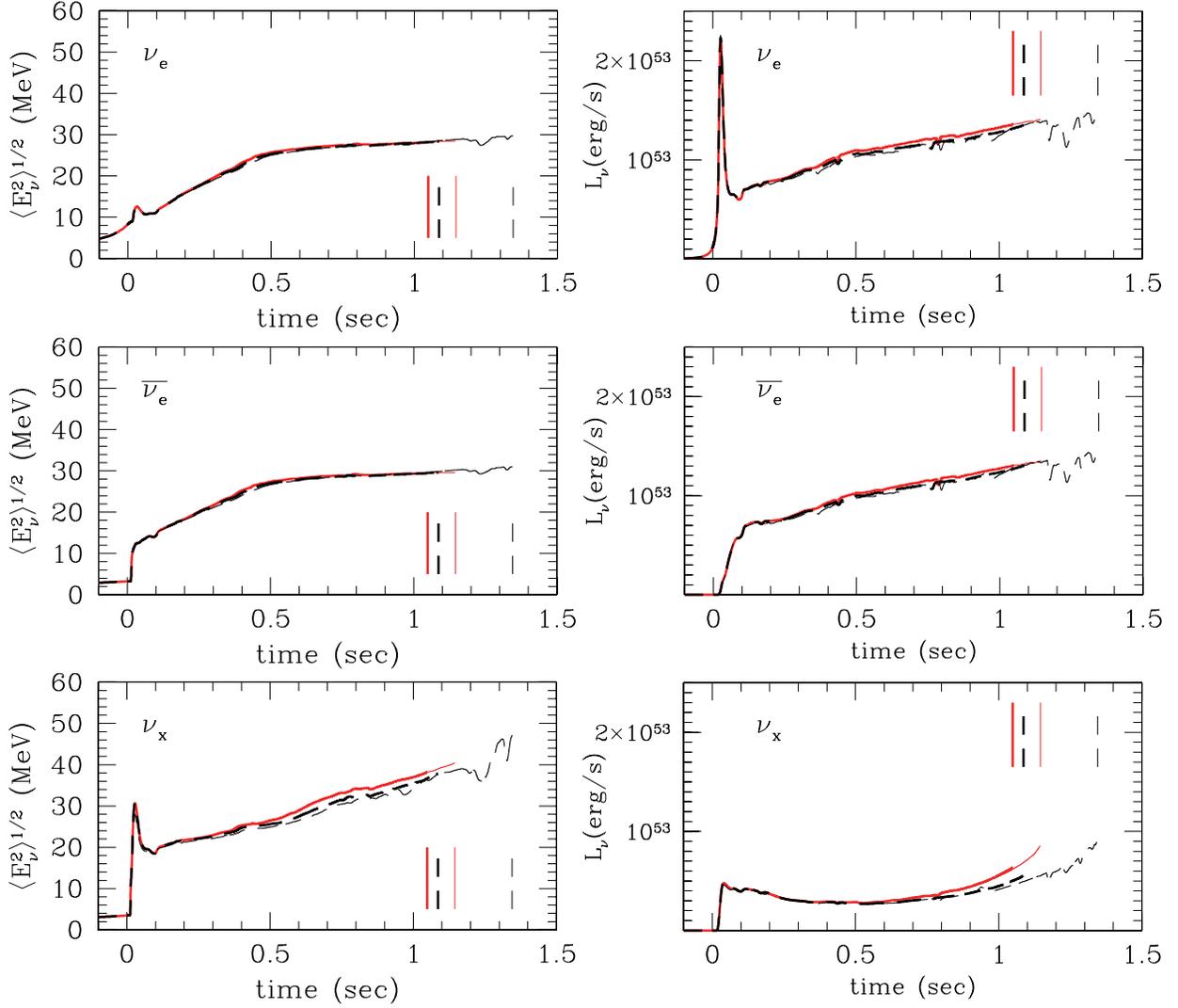}
\caption{Average energies (left) and luminosities (right) of neutrinos emitted from $40M_\odot$ models as a function of time after bounce. The panels correspond, from top to bottom, to $\nu_e$, $\bar\nu_e$ and $\nu_x$ ($=\nu_\mu$, $\nu_\tau$, $\bar\nu_\mu$, $\bar\nu_\tau$). Vertical lines represent the end point of the neutrino emission. The notation of lines is the same as that in Figure~\ref{cd40}.}
\label{neuprof}
\end{figure}

\begin{figure}
\plotone{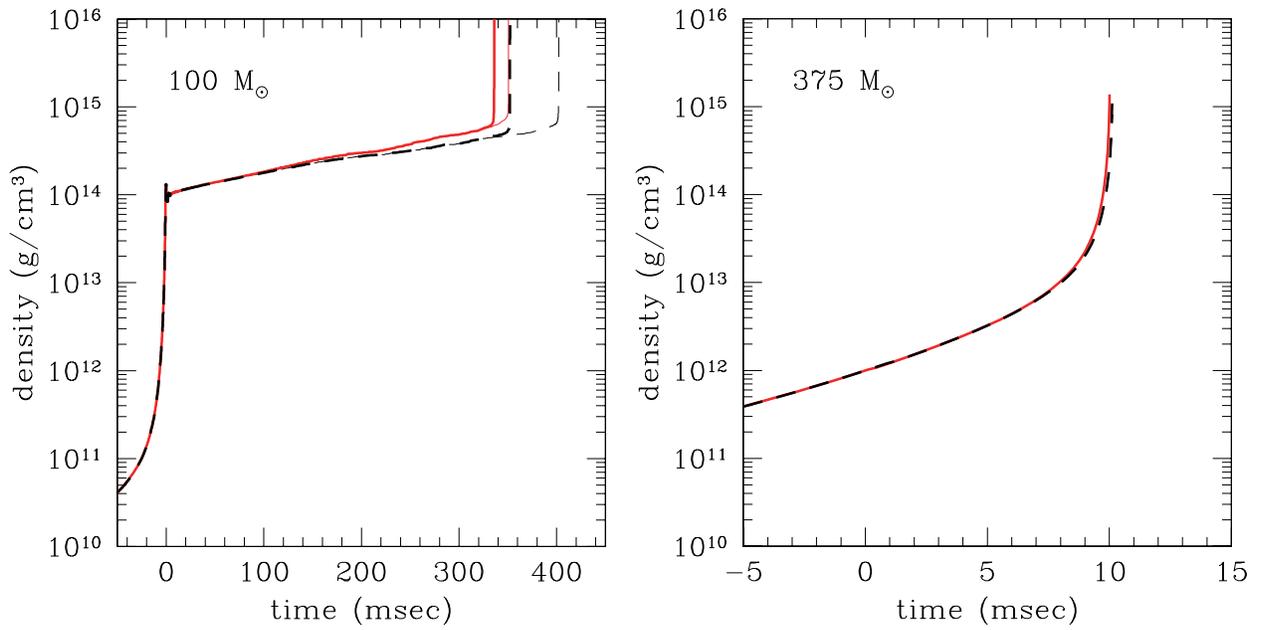}
\caption{Same as Figure~\ref{cd40} but for the models with $100M_\odot$ (left) and $375M_\odot$ (right). Note that the time is measured from the point when the central density exceeds $10^{12}$~g~cm$^{-3}$ for the models with $375M_\odot$, whereas it is measured from the bounce for the models with $100M_\odot$ and $40M_\odot$ (Figure~\ref{cd40}).}
\label{cdmd}
\end{figure}

\begin{figure}
\plotone{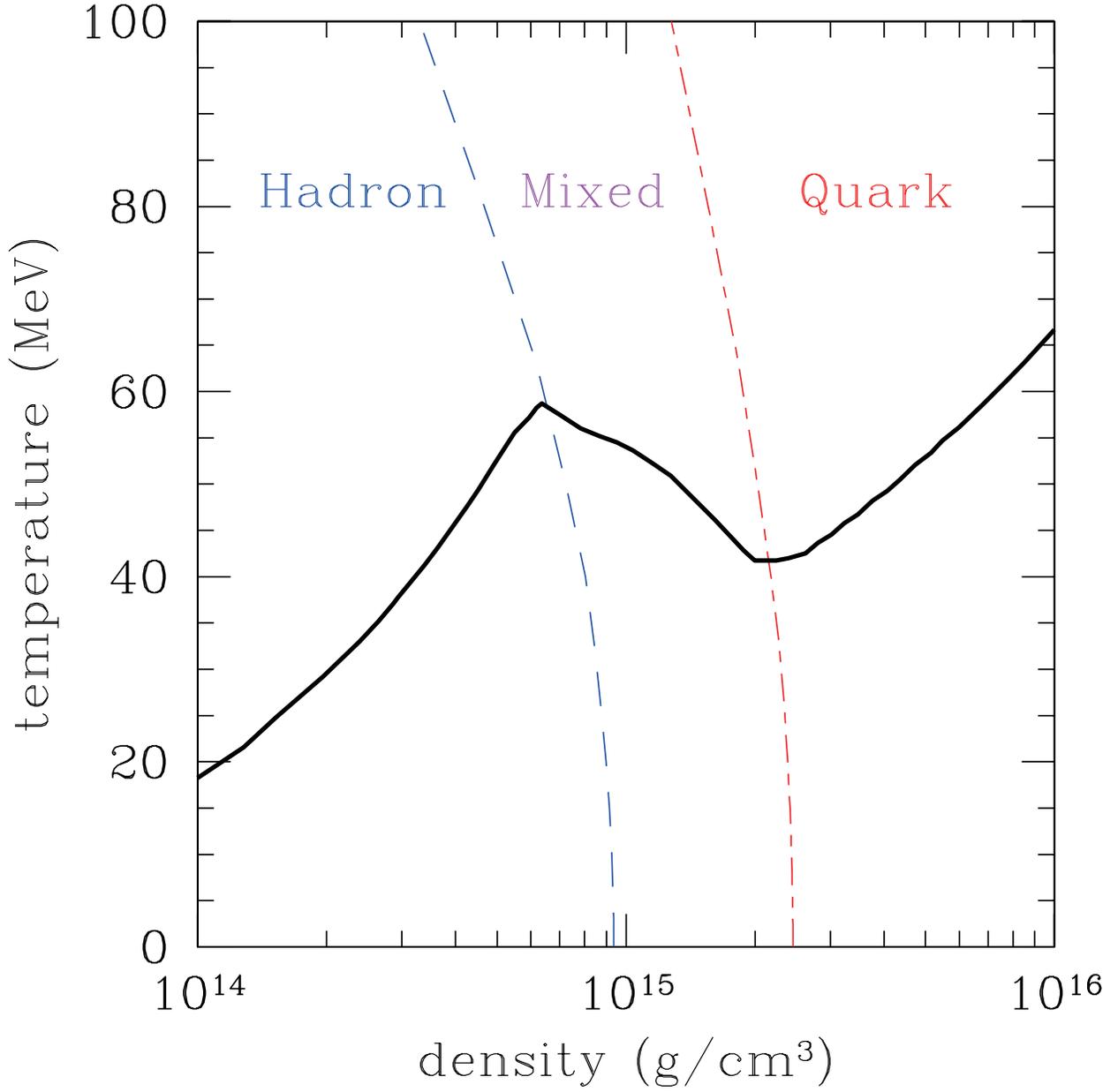}
\caption{Same as Figure~\ref{dent-ph} but for the evolution of $100M_{\odot}$ model with EOS PQ and the phase diagram for $Y_l=0.17$.}
\label{dent-ph100}
\end{figure}

\begin{figure}
\plotone{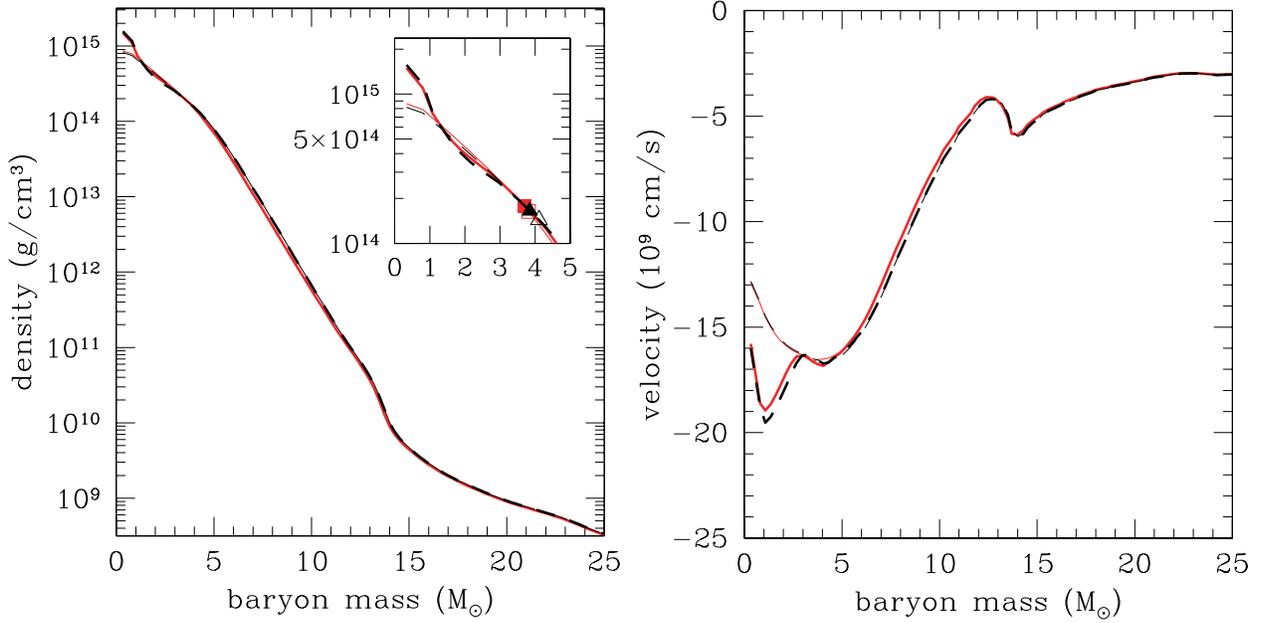}
\caption{Profiles of the density (left) and radial velocity (right) for the models with $375M_\odot$ at the time of apparent horizon formation. The notation of lines is the same as that in Figure~\ref{cd40}. In the close-up plots in the left panel, the empty triangle, filled triangle, empty square and filled square, show the locations of the apparent horizon for the models with EOS's~OO, OQ, PO and PQ, respectively.}
\label{dv375}
\end{figure}

\begin{figure}
\plotone{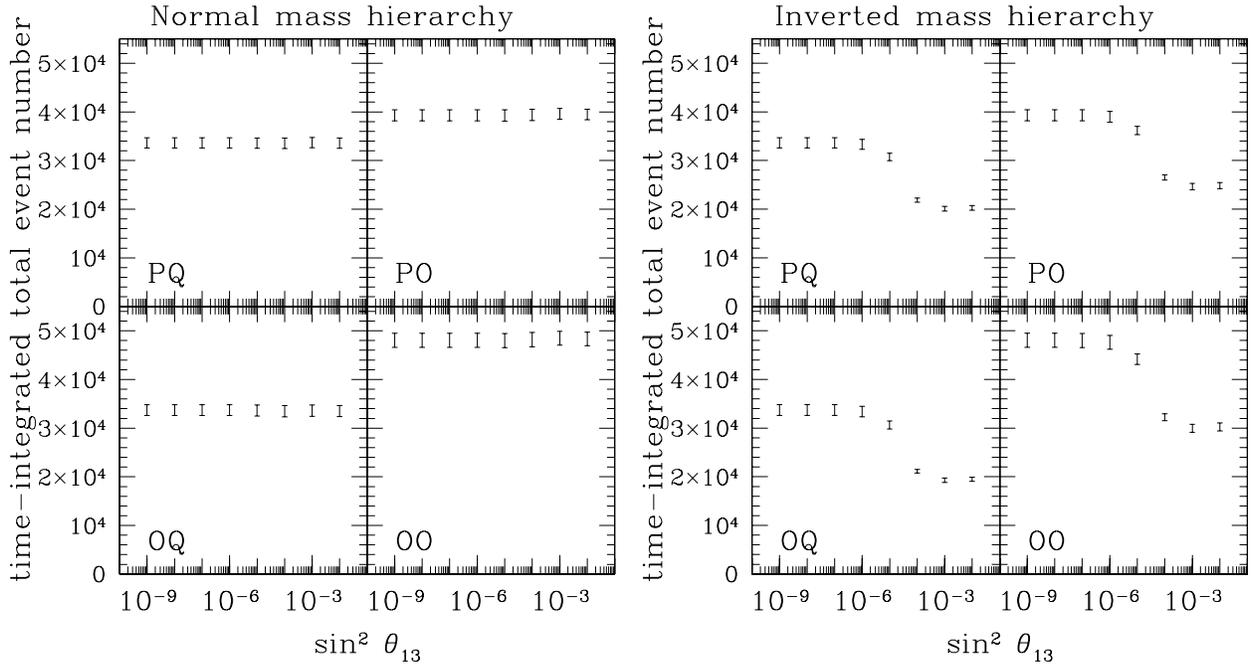}
\caption{Time-integrated neutrino event numbers of $40M_\odot$ models for the normal mass hierarchy (left) and the inverted mass hierarchy (right). Error bars represent the upper and lower limits owing to different nadir angles. In each panel, the upper left, upper right, lower left and lower right plots correspond to the models with EOS's~PQ, PO, OQ, and OO, respectively.}
\label{ndettot}
\end{figure}

\begin{figure}
\plotone{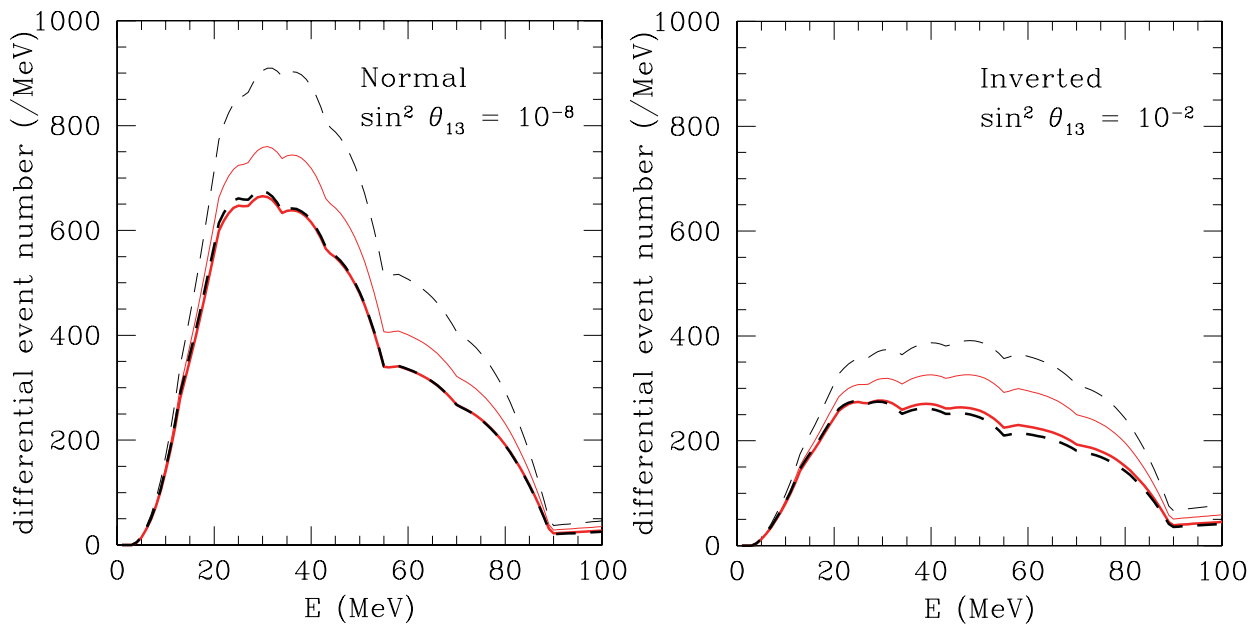}
\caption{Energy spectra for the time-integrated event number of neutrinos in the case without the earth effects. The left and right panels correspond to the cases of the normal mass hierarchy with $\sin^2\theta_{13}=10^{-8}$ and the inverted mass hierarchy with $\sin^2\theta_{13}=10^{-2}$, respectively. The notation of lines is the same as that in Figure~\ref{cd40}.}
\label{ndetspc}
\end{figure}

\end{document}